\documentclass{sig-alternate-2013}

\usepackage[english]{babel}
\usepackage[utf8x]{inputenc}
\usepackage[T1]{fontenc}
\usepackage[font={small,it}]{caption}

\usepackage[show]{chato-notes} % to show notes.

\usepackage{amsmath}
\usepackage{graphicx}
\usepackage[colorlinks=true, allcolors=blue]{hyperref}
\newcommand{\spara}[1]{\smallskip\noindent\textbf{#1}}

\makeatletter %%%%% REMOVES ACM COPYRIGHT BOX. REMOVE FOR FINAL VERSION
\def\@copyrightspace{\relax}
\makeatother

\title{Understanding International Migration using Tensor Factorization\thanks{Accepted as poster at WWW 2017}}
\author{
\end{tabular}\begin{tabular}[t]{p{0.4\textwidth}}\centering
Hieu Nguyen\\
      \affaddr{Aalto University}\\
      \affaddr{Espoo, Finland} \\
      \texttt{hieu.nguyen@aalto.fi}
\end{tabular}\begin{tabular}[t]{p{0.4\textwidth}}\centering
Kiran Garimella\\
       \affaddr{Aalto University}\\
       \affaddr{Espoo, Finland}\\
       \texttt{kiran.garimella@aalto.fi}
}

\begin{document}
\maketitle

%\begin{abstract}
\section{Introduction}
Understanding human migration is of great interest to demographers and social scientists. 
User generated digital data has made it easier to study such patterns at a global scale. Geo coded Twitter data, in particular,  has been shown to be a promising source to analyse large scale human migration. 
But given the scale of these datasets, a lot of manual effort has to be put into processing and getting actionable insights from this data. 

In this paper, we explore feasibility of using a new tool, tensor decomposition, to understand trends in global human migration. 
We model human migration as a three mode tensor, consisting of (origin country, destination country, time of migration) and apply CP decomposition to get meaningful low dimensional factors.
Our experiments on a large Twitter dataset spanning 5 years and over 100M tweets show that we can extract meaningful migration patterns.% with out much manual effort.

\section{Related Work} \label{relatedwork}
Understanding human mobility patterns at a global scale using digital data has been of great interest to demographers and social science researchers.
%~\cite{hughes2016inferring}.
In particular, geo-tagged Twitter data has been used extensively in the past to study global human mobility patterns~\cite{hawelka2014,zagheni2014inferring}.
%An extensive survey on the traditional and modern approaches to analyse human migration patterns has been summarized in \cite{hughes2016inferring}.
Tensors are higher dimensional extensions of matrices, which can be used to represent multi modal data. A comprehensive survey on tensors and applications of tensors can be found in~\cite{kolda2009tensor}.
Tensor factorization provides a principled way to analyse large scale multi-modal datasets. Recent progress on scalable implementations of tensor factorization~\cite{schein2015} have lead to the application of tensors in a wide range of fields, including Criminology, Neuroscience, Socialscience, etc. See~\cite{fanaee2016tensor} for a detailed survey.
Our paper complements existing work on using Twitter data by showing the applicability of a new tool (tensor factorization) to better understand large scale migration behavior.

%analysis was used to model Foursquare location check-in dataset and extract interesting spatio-temporal patterns in the work of \cite{papalexakis2015location}. BPTF was introduced by \cite{schein2015} and was used to analyse massive dyadic event dataset. We follow their framework and apply it to our dataset.

%\cite{wen2016pairfac} ..

%to the best of our knowledge, we are the first to apply tensor decomposition to understand large scale migration.

\section{Data} \label{dataset}

%\subsection{Collecting data and pre-processing}
Using the Archive Twitter stream\footnote{\small\url{https://archive.org/details/twitterstream}} (1\% random sample) from 2011--2016, we obtained 138M geo-tagged tweets. We used the geo coordinates (lat,long) to obtain the country from which a user tweeted and filtered out users who had a geo tagged tweet in at least two countries from October 2011 to November 2016 giving us 428,000 users. Using the Twitter API, we obtained the 3,200 most recent tweets for all these 428k users, which gave us 109M geo tagged tweets.
%Each tweet has several attributes such as the lattitude, the longitude and the tweeting time. From the tweets' coordinates, we extract the country which the tweets belong to.
%
We preprocessed the data using simple heuristics, used in previous work~\cite{hawelka2014} to remove noise and bot accounts.
We defined a user's monthly country of residence as the country where she produces most of the tweets in that month. If a user doesn't tweet at all in a month, we assign that month with the most recently known country of residence. 
%If the user tweets from multiple countries, we take the country from which a majority of the tweets appear.
%For example, if a user doesn't tweet from April to September and he last tweeted in March in USA, we will set the country of residence from April to September as USA. 
We define a migration at some month as a change of country of residence between windows of $k$ months before and after that month. 
This way, by simply adjusting the window size, we can analyse the migration patterns of different types, such as short-term migration (e.g. student's one semester exchange) or long-term migration (e.g. permanent migration).
%For example, considering a 3-month window, if a user resides in the US in March and May and in Canada in April, we count that she resides in US from March to May. Next, she resides in Germany from June to August, we count he has migrated from US to Germany in June. 

%Before we perform the analysis, we preprocess the data to remove the artificial and the network noise, which may effect the results. Because Twitter is a popular social network, there are many social bots due to the economic and the political incentives. Our approach is purely heuristic and follows multiple rules of thumbs shown in the paper \cite{ferrara2016} and \cite{hawelka2014}. Firstly, we assumed the twitter social bots have certain properties such as their the number of followers less than 10, their number of friends less than 10, their account creation date less than one years. We remove the tweets from these accounts. Secondly, we remove the network noise by discarding consecutive tweets which have the location changing speed more than 1000 km/h. After the pre-processing steps, about 96.2\% of the users and 97.6\% of the tweets remain in our dataset.

%\subsection{Definition of country of residence and migration}

\section{Methodology} \label{methodology}

%\subsection{Tensor modeling}

An $n$-way tensor is a generalization of a matrix (2-way tensor). After getting the migration history of the users, we aggregate the global migration flow as an 3-way tensor $A$ with the size $N \times N \times M$, where $N=228$ is the number of countries and dependent territories in our dataset and $M=74$ is the number of time-steps (monthly from October 2010 to November 2016). The entry $A[i, j, k]$ is the number of Twitter users migrating from the country $i$ to the country $j$ at the time-step $k$. 
%In the 1-month, 3-month and 5-month tensor window, the percentage of nonzero entries are about 4.4\%, 0.0084\% and 0.51\% correspondingly.

%\subsection{Exploratory analysis using tensor factorization}

A standard technique to decompose a matrix into its salient components (factors) is Singular Vector Decomposition (SVD). For tensors, a generalization of SVD, called CP decomposition~\cite{kolda2009tensor}, can be used to obtain the salient factors.
%To do that on a tensor, we can use a technique called Canonical Polyadic (CP) or PARAFAC, which is a generalization of SVD. 
Suppose $A$ is a 3-way tensor and $K$ is a positive integer. A CP decomposition decomposes $A$ into three latent $factor\ matrices$, which are a sum of $K$ component rank-one tensors.
%a sum of $K$ rank-one components as follows:
%
\begin{align}
  A \approx [O, D, T] =\sum_{i=1}^K o_i \circ d_i \circ t_i
\end{align}
i.e., the tensor can be represented as the sum of $K$ components of the outer product of three vectors. 
Each vector ($o,d,t$) corresponds to one of the three dimensions of the tensor. 
%For example, in our setting where we model the migration data as a 3-way tensor, 
%$A$ (origin country, destination country, time-step) with the entry as the count. 
Vector $o_i$ represents a factor corresponding to the origin country, $d_i$ the destination country and $t_i$ the time-step. For each of three dimensions, we can stack $K$ vectors (components) as $K$ columns of a matrix, which is called a $factor\ matrix$. 
In our case, we have three factor matrices $O$, $D$ and $T$ which have the size $N \times K$, $N \times K$ and $M \times K$ respectively.
Each factor matrix is a $K$-dimensional representation of the salient patterns in the migration counts.
In this paper, we used Bayesian Poisson Tensor Factorization (BPTF)~\cite{schein2015} for the CP decomposition, since it handles sparse tensors effectively.

%Because Twitter data is under-representative in certain regions and the human migration between certain pairs of countries are low, the count tensor is often sparse. Traditional tensor decomposition techniques may not perform well when fitting sparse tensor \cite{chi2012tensors}. Bayesian Poisson Tensor Decomposition (BPTF) was introduced in \cite{schein2015} and was shown to perform better than traditional models on sparse and disperse tensor. We use their software to perform tensor decomposition on our data with the default parameters because it runs fast. Experimenting thoroughly with different parameter values for BPTF is beyond the scope of our study.

To obtain interesting insights from the factor matrices, we start with measuring the distribution of the components in the time-step factor ($T$).
The components having many uneven values in the time-step vector $t_i$ may represent interesting patterns such as sudden spikes in migration. 
To measure the uneven distribution in the time component, we compute the Gini Coefficient\footnote{\small{\url{https://en.wikipedia.org/wiki/Gini_coefficient}}} of each of the $K$ components,
%After decomposing the tensor into factor matrices, we measure the sparsity of the time-step factors in the matrix T using the Gini coefficient \cite{schein2015}. 
and rank the top-10 components with highest Gini values along with the corresponding origin country and destination country factors $o_i$ and $d_i$. 
In this way, we can analyse the top origin and destination countries countributing abnormally in the time-step factor. 
After that, we plot and examine the most deviant components in the order of the Gini coefficient rank. 
%The higher the Gini coefficient, the more likely the time-step vector has some interesting pattern.

\section{Findings} \label{findings}
\begin{figure}[h]
  \centering
  \includegraphics[width=\columnwidth, height=.2\textheight, clip=true, trim=0 15 0 0]{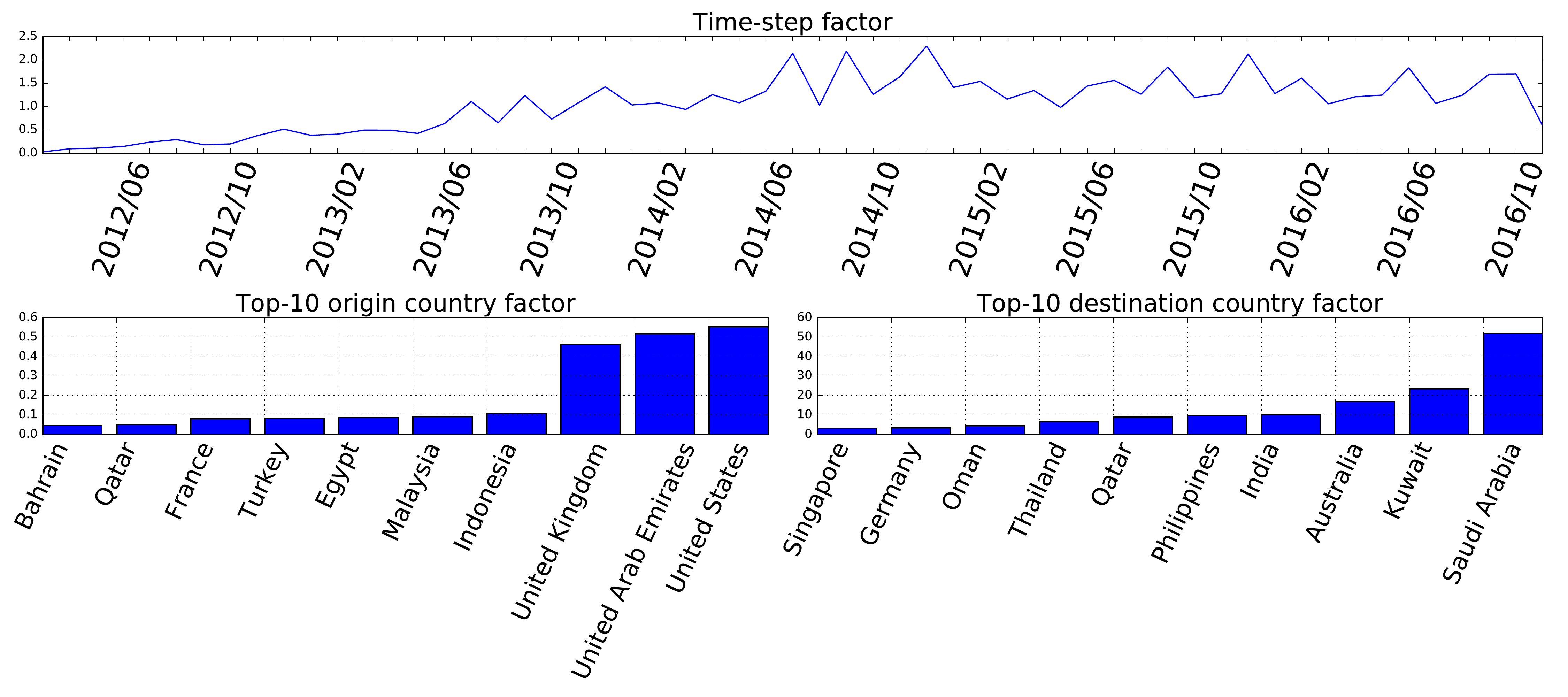}
  \caption{Components extracted from setting $k$ to 1-month. The top origin countries are UK, UAE and USA. The top destination countries are Kuwait and Saudi Arabia. From 2013 to 2016, we noticed that there is a yearly peak in the timestep factor usually in December. Because the 1-month window favors counting visitors' short trips, we hypothesize this pattern represents tourist travel.}
  \label{fig:month1}
\vspace{-\baselineskip}
\end{figure}

\begin{figure}[h]
  \centering
  \includegraphics[width=\columnwidth, height=.2\textheight, clip=true, trim=0 15 0 0]{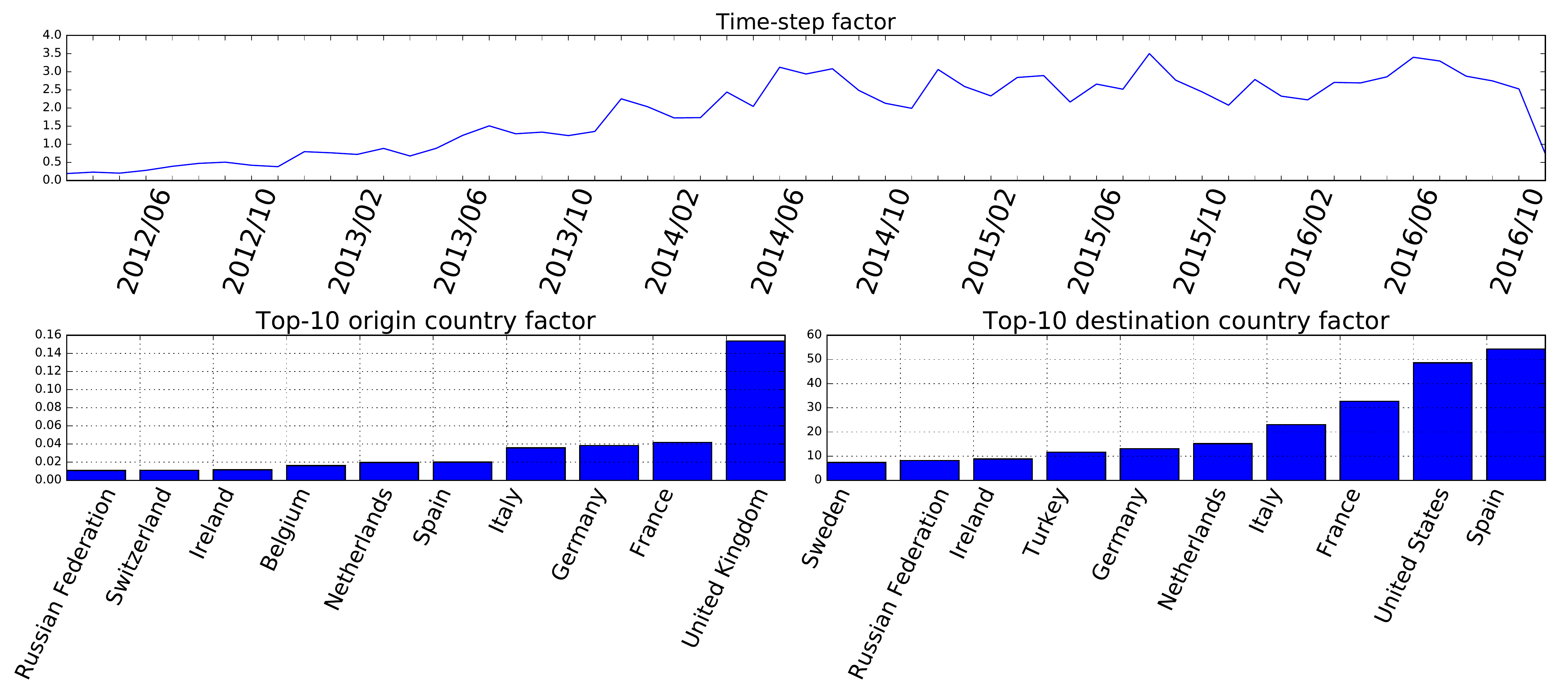}
  \caption{Components extracted from setting $k$ to 3-months. The top origin and destination countries are UK and Spain correspondingly. The other top countries are also from Europe and USA. From 2014 to 2015, we notice the high peaks in around August and the smaller peaks in around December. We don't consider 2016 because the dataset only has part of November 2016 and no December 2016. Our hypothesis is that this pattern may represent European student's Autumn study exchange, which typically lasts 3 months.}
  \label{fig:month3}
\vspace{-\baselineskip}
\end{figure}

\begin{figure}[h]
  \centering
  \includegraphics[width=\columnwidth, height=.2\textheight, clip=true, trim=0 15 0 0]{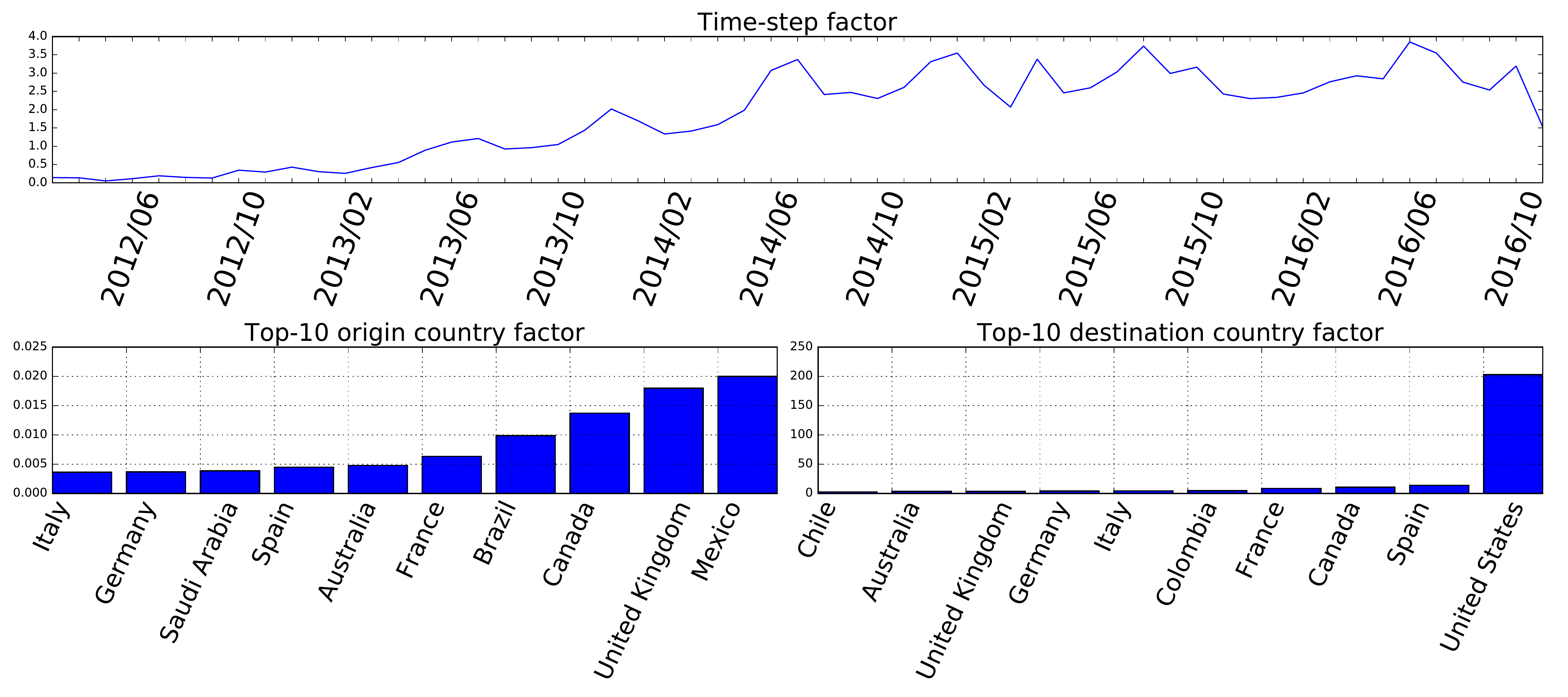}
  \caption{Components extracted from setting $k$ to 5-months. The top origin countries are Mexico, UK and Canada. The top destination country is the US. Our hypothesis is that this pattern may represent the migration flow to USA (for working or for permanent residency).}
  \label{fig:month5}
\vspace{-\baselineskip}
\end{figure}

Using the above methodology, we constructed a 3-mode tensor with $k$ = 1, 2, 3, 4 and 5 month windows, with low rank $K$ = 15 components and examined the results. We find some interesting observations.
(i) Setting a low $k$, say, 1 month, we are able to capture events related to tourist migration, see Figure~\ref{fig:month1}, (ii) setting $k$ to around 3 months, we find patterns related to Erasmus student migration around Europe, as seen in Figure~\ref{fig:month3}, and (iii) setting $k$ to 5 or more, we find patterns related to long term migration, see Figure~\ref{fig:month5}.\footnote{\small{Due to lack of space, we only show one component. The remaining components also contain meaningful information, and can be seen here \url{https://www.dropbox.com/sh/9jyxddxzrd4kcwb/AADXvcBHMKk_HSosOyoUYknBa?dl=0}}.} 

%Next, we decomposed the tensors to 15 and 50 components. Next, we plot the graphs as described in the section \ref{methodology} and examine them carefully. We try to interpret the patterns with our assumption and hypothesis. Also, we perform the search with keywords: top origin country, top destination country and time on Google and Wikipedia to check the global events that we may not know. We show two components' plots with two different window size in the figure \ref{fig:month3} and \ref{fig:month5}, which may represent meaningful events.

\spara{Discussion} Our paper shows the potential of the application of tensor decomposition methods to get insights from large scale human migration on Twitter. Our results show that this could be a useful tool to summarizing large volumes of complex interactions, which can be inspected by domain experts to take further action.
We restricted ourself to three modes, for simplicity of presentation. We can easily incorporate more modes, like topics being discussed by the users tweets, to get an understanding on what the migrating users speak about.
%Finally, our paper is meant to provide insights that can be inspected by demographers to take further action. 
%Note that we do not claim to be a one stop solution to understand human migration. 
%Our intent with this paper is to provide a way to better understand complex interactions between the various modes. 
%Though we show that...
%Finally, the
%The analysis results might be biased towards the regions having high Twitter penetration rate. To correct the bias, additional processing steps may be needed %as stated in \cite{hughes2016}. 
%The visualization is available at \cite{vizlink}.

%\section{Conclusions} \label{conclusions}

%Some conclusions.
%\subsection{Limitations of Twitter data}
%The analysis results might be biased towards the regions having high Twitter penetration rate. To correct the bias, additional processing steps may be needed %as stated in \cite{hughes2016}. 
%However, obtaining a bias-free dataset is beyond the scope of this study. We focus on applying the tensor analysis techniques to extract meaningful patterns and assume that our results are only relevant within the Twitter data context. 
%To visualize the non-representative regions in our dataset (such as Africa and China), we use the method and code by \cite{abel2014}, which is inspired by \cite{krzywinski2009}. 

\bibliographystyle{abbrv}
\bibliography{related}

\end{document}